\documentclass[aps,pre,twocolumn,showpacs,superscriptaddress,groupedaddress]{revtex4}  
\usepackage{graphicx}  
\usepackage{dcolumn}   
\usepackage{bm}        
\usepackage{amssymb}   
\usepackage{amsmath}
\usepackage{xcolor}
\usepackage{amsthm}
\usepackage{enumitem}
\usepackage{upgreek}

\usepackage{dsfont}
\newtheorem{theorem}{Theorem}

\begin{document}


\title{When Shannon and Khinchin meet Shore and Johnson: equivalence of information theory and statistical inference axiomatics}

\author{Petr Jizba}
\email{p.jizba@fjfi.cvut.cz}
\affiliation{Faculty of Nuclear Sciences and Physical Engineering,
Czech Technical University in Prague, B\v{r}ehov\'{a} 7, 115 19, Prague, Czech Republic}

\author{Jan Korbel}
\email{jan.korbel@meduniwien.ac.at}
\affiliation{Section for Science of Complex Systems, Medical University of Vienna, Spitalgasse 23, 1090 Vienna, Austria}
\affiliation{Complexity Science Hub Vienna, Josefst\"{a}dter Strasse 39, 1080 Vienna, Austria}
\affiliation{Faculty of Nuclear Sciences and Physical Engineering,
Czech Technical University in Prague, B\v{r}ehov\'{a} 7, 115 19, Prague, Czech Republic}

\begin{abstract}
We propose a unified framework for both Shannon--Khinchin and Shore--Johnson axiomatic systems.
We do it by rephrasing Shannon--Khinchine axioms in terms of generalized arithmetics of Kolmogorov and Nagumo.
We prove that the two axiomatic schemes yield identical classes of entropic functionals --- the Uffink class of entropies.
This allows to re-establish the entropic parallelism between information theory and statistical inference
that has seemed to be ``broken'' by the use of non-Shannonian entropies.
\end{abstract}

\vspace{-1mm}
\maketitle
%

\section{Introduction}
Entropy is undoubtedly one of the most important concepts in physics, information theory and statistics~\cite{threefaces}. The notion of entropy was originally developed by Claussius,
Boltzmann, Gibbs, Carath\'{e}odory and others in the context of
statistical thermodynamics.
There it supplemented a new state function that was naturally extensive
(due to its very formulation in terms of the heat 1-form) and  in any adiabatically isolated system it represented a non-decreasing function of its state variables (on account of the Clausius theorem).
Roughly a half-century after these developments, the entropy paradigm was further conceptualized in the theory of information by Shannon~\cite{shannon}. In this later context the ensuing
entropy (Shannon's entropy or measure of information) quantitatively represented the minimal number of binary (yes/no) questions which brings us from our present state of knowledge about the system in question
to the one of certainty. The higher is the measure of information (more
questions to be asked) the higher is the ignorance about the system and thus
more information will be uncovered after an actual measurement. A proper axiomatization of
Shannon's entropy is encapsulated in the so-called Shannon--Khinchin (SK) axioms~\cite{khinchin}.
Only one decades after Shannon's seminal works, Jaynes~\cite{jaynes1,jaynes2} promoted Shannon's information measure to the
level of inference functional that was able to extract least biased probability distributions from measured data. This procedure is better known as \emph{Maximum entropy principle} (MEP).
Since MEP is, in its essence, a statistical inference method, it needs a proper mathematical qualification to place Jaynes'  heuristic arguments in a sound mathematical framework. The corresponding mathematical qualification was provided by Shore and Johnson (SJ) in the form of axioms that ensure that the MEP estimation procedure is consistent with desired properties of inference methods~\cite{shore,shoreII}. At this point, one should emphasize that in the statistical inference theory (SIT) entropy functionals serve only as convenient technical vehicles for unbiased assignment of  distributions that are compatible with given constraints.  In fact, one might say that it is the MEP distribution that is the primary object in SIT while the entropy itself is merely secondary (not having any operational role in the scheme).
This is very different from the information theory or thermodynamics where entropies are primary objects with firm operational meanings (given, e.g., in terms of coding theorems or calorimetric measurements). In the original paper~\cite{shore,shoreII} Shore and Johnson concluded that
their axioms yield only one ``measure of bias'', namely Shannon entropy.
It might, however, seem a bit puzzling why ``measure of bias'' should have anything to do with additivity (i.e., one of the defining properties of Shannon's entropy).
In the end, any monotonic function of such a measure should provide the same MEP distribution but might (and as a rule it does) yield  non-additive entropy.
So, it is perhaps not so surprising that with the advent of generalized entropies~\cite{jizba3,tsallis2,havrda,sharma,Kaniadakis,hanel,arimitsu,korbel,wada}, the past two decade have
seen a renewed interest both in the SJ axiomatics and the associated classes of admissible entropies~\cite{presse,tsallis,presse2,bagci,jizba19,uffink}.
In particular, it has been shown in Ref.~\cite{jizba19}  that the SJ axiomatization of the inference
rule does account for substantially wider class of entropic functionals than just SE  --- the so-called Uffink class~\cite{uffink}, which include Shannon's entropy as a special case.


The main aim of this paper is to answer the following question: what generalization of the SK axioms would provide the Uffink class of entropic functional?  This would not only allow to re-establish the ``broken'' entropic parallelism between information theory and statistical inference but it should also cast a new light on the Uffink class of entropies and its practical utility.

We first recall the original set of SK axioms~\cite{khinchin}:

Let $A$ and $B$ be two discrete random variables  with respective sets of possible values ${\mathcal{A}} = \{A_i\}_{i=1}^n$ and ${\mathcal{B}} = \{B_i\}_{i=1}^m$.  With ${\mathcal{A}}$ one can associate a complete set of events $\{a_i\}_{i=1}^n$ so that $a_i$ denotes the event that $A=A_i$ (similarly for ${\mathcal{B}}$). Elements $a_i$ (and $b_j$) are known as elementary events.
%
%
Let
\begin{eqnarray*}
&&P(A = A_i) = P(A_i) = p_i, \;\; P(B_j) = q_j\, , \;\;\; \nonumber \\[1mm]
&&P(A = A_i, B = B_j) = P(A_i, B_j) = r_{ij}\, , \nonumber \\[1mm]
&&P(A = A_i|B = B_j) = P(A_i|B_j) = r_{i|j} = r_{ij}/q_j\, ,\nonumber \\[1mm]
&&1~\leq~i~\leq~m;\;\; 1~\leq~j~\leq~n\, ,
\end{eqnarray*}
be corresponding elementary-event, joint and conditional probabilities, respectively.
For $A$ and $B$ we denote the ensuing probability distributions as $P_A = \{p_i\}_{i=1}^n$ and $P_B = \{q_j\}_{j=1}^m$. Likewise, we write $P_{A,B} = \{r_{ij}\}_{i,j =1}^{n,m}$,  $P_{A|B} = \{r_{i|j}\}_{i,j =1}^{n,m}$
and $P_{A|B_j} =  \{r_{i|j}\}_{i=1}^{n}$.
The entropy of the probability distribution $P_A$  (which may also be called the entropy of $A$) will be, with a slight abuse of the notation, denoted interchangeably as ${\mathcal{H}}(P_A)$ or ${\mathcal{H}}(A)$. Similar notation  will be introduced for distributions $P_B$, $P_{A,B}$,$P_{A|B}$ and $P_{A|B_j}$.

%
%
\begin{enumerate}[label = {\bf SK{\arabic*}}]
\item \emph{Continuity:} Entropy is a continuous function w.r.t. all its arguments, i.e.,
$\mathcal{H}(P) \in \mathcal{C}.$
\item \emph{Maximality:} Entropy is maximal for uniform distribution, i.e.,
$\max_P \mathcal{H}(P) = \mathcal{H}(U_n)$,
where $U_n = \{1/n,\dots,1/n\}$.
\vspace{1mm}
\item \emph{Expandability:} Adding an elementary event with probability zero does not change the entropy, i.e,
\vspace{-2mm}
$$\mathcal{H}(p_1,\dots,p_n,0) = \mathcal{H}(p_1,\dots,p_n).$$
\vspace{-8mm}
%
\end{enumerate}
\begin{description}
\item[{\bf SK4$^S$}] \emph{Shannon additivity:}
\vspace{-2mm}
$$~~~~\mathcal{H}(A,B) = \mathcal{H}(A|B)+\mathcal{H}(B) = \mathcal{H}(B|A) + \mathcal{H}(A),$$
where $\mathcal{H}(B|A) = \sum_i p_i \ \! \mathcal{H}(B|A=A_i)$.
\end{description}
We note that the conditional entropy $\mathcal{H}(B|A)$ can be calculated in two ways:
{\em{i)}} from the entropy of the joint distribution of the pair $(A,B)$ and marginal distribution of $A$,
or {\em{ii)}} from the marginal distribution $A$ and entropy of the conditional random variable $B|A\!=\!A_i$.
This duality is crucial for the internal consistency the SK axiomatic scheme. Aforestated set of SK axioms has the unique solution --- Shannon's entropy~\footnote{Here and throughout we use the base of natural logarithms.
Entropy thus defined is then measured in natural units --- {\em nats}, rather than {\em bits}.
To convert, note that 1 bit = 0.693 nats.}
\begin{equation*}
\mathcal{H}(P) \ = \ - \sum_i p_i \log p_i\, .
\end{equation*}

With the advent of generalized entropies~\cite{jizba3,tsallis2,havrda,sharma,Kaniadakis,hanel,arimitsu,korbel,wada} there arose two natural questions. First, is it possible to conceptualize such entropies in terms of information-theoretic axioms ({\em\`{a}~la} SK axioms)? And second, can generalized entropies be used as consistent inference functionals with sound mathematical underpinning ({\em\`{a}~la} SJ axioms)? As for the first question, it is well known that one can ``judiciously'' generalize
the additivity axiom {\bf SK4$^S$} to produce various generalized entropies. Typical examples are provided by R\'{e}nyi and Tsallis--Havrda--Charv\'{a}t (THC) entropies. For instance, for the R\'{e}nyi entropy, one keeps axioms {\bf SK1-3} and substitute {\bf SK4$^S$} with~\cite{jizba3}:
\begin{description}
\item[{\bf SK4}$^R$]
\emph{R\'{e}nyi additivity:} $\mathcal{R}_q(A,B) =  \mathcal{R}_q(A|B) \ +\ \mathcal{R}_q(B)$\\  $~~= \ \mathcal{R}_q(B|A) \ + \ \mathcal{R}_q(A)$, where $\mathcal{R}_q(B|A)\ =$\\ \mbox{\hspace{2mm}$f^{-1}\!\left(\sum_i \rho_i^A(q) f\left(\mathcal{R}_q(B|A=A_i)\right)\right)$}.
\end{description}
Here, $\rho_i^A(q) = {(p_i)^q}/{\sum_j (p_j)^q}$ is the escort (or zooming) distribution~\cite{beck,Mendes} and $f$ is an arbitrary invertible and positive function on $[0, \infty)$.
Corresponding axiomatics is stringent enough to fix uniquely $f(x)$ to be
either $f(x) = e^{(1-q)x}$ (for $q\neq 1$) or $f(x) = x$ (for $q=1$), and  yields the R\'{e}nyi entropy
\begin{equation*}
\mathcal{R}_q(P) \ = \ \frac{\log \sum_i p_i^q}{1-q}\, ,
\end{equation*}
as the unique solution.

Similarly, for the case of non-additive THC entropy~\cite{tsallis2,havrda}
one can augment axioms {\bf SK1-3} with~\cite{abe,suyari}
\begin{description}
  \item[{\bf SK4}$^T$]  \emph{Tsallis additivity:}  $\mathcal{S}_q(A,B)  =  \mathcal{S}_q(B|A) + \mathcal{S}_q(A)$\\
      \mbox{$~~+ (1-q)  \mathcal{S}_q(B|A) \mathcal{S}_q(A)$}
where $\mathcal{S}_q(B|A) =$ \\$~~~\  \sum_i \rho_i^A(q) \mathcal{S}_q(B|A=A_i)$,
\end{description}
where $\rho_i^A(q)$ is again the escort distribution.
The unique solution of this axiomatic system gives the THC entropy
\begin{equation*}
\mathcal{S}_q(P)  \ = \  \frac{\sum_i p_i^q-1}{1-q}\, .
\end{equation*}

In parallel with this there has been several successful attempts to classify  entropic functionals according to various desirable information-theoretic properties. Here we should mention, e.g., the class of strongly pseudo-additive entropies (SPA) based on generalization of R\'{e}nyi entropy axioms for non-additive entropies~\cite{ilic}, $Z$-entropies based on group properties of the entropic functionals~\cite{tempesta} or classification according to the asymptotic scaling leading to (c,d)-entropies~\cite{hanel} and ensuing generalizations~\cite{korbel18}.

As for the second question, there has been notable progress in recent years in the classification of entropic functionals satisfying
SJ axioms~\cite{uffink,jizba19,jizba19b}. Our aim here is to employ generic arithmetical principles to generalize, in a logically sound way, the SK  axiomatic scheme.  To this end we will use the framework of Kolmogorov--Nagumo (KN)
arithmetics~\cite{Czachor_II,Naudts}, KN quasi-arithmetic means~\cite{kolmogorov,nagumo,Czachor} and escort distributions~\cite{beck,Mendes}. Ensuing class of admissible entropies will be compared with the class of entropies solving SJ axioms --- Uffink class. We will see that
both classes not only coincide, and hence bolster the entropic parallelism between information theory and statistical inference, but there also is a close parallelism between the two axiomatic schemes.

The rest of the paper is organized as follows: In Section~II, we briefly summarize the concept of generalized arithmetics and outline the key
role that Kolmogorov--Nagumo functions play in this context. In Section~III, we introduce the class of Shannon--Khinchin axioms based on the
Kolmogorov--Nagumo generalized arithmetics and derive the generic class of entropic functionals satisfying these axioms. In Section~IV, we show
the equivalence of the aforementioned class and the Uffink's entropic class. This will, in turn, cast new light on the relationship between SK and SJ axiomatic schemes.
This will in turn re-establish the entropic parallelism between information theory and statistical inference.
The last section is devoted to some further observations, remarks and conclusions.

\section{Generalized arithmetics and Kolmogorov and Nagumo functions}

Let us now introduce the concept of generalized arithmetics.
From abstract algebra it is known that arithmetic operations can be defined in various ways, even if one assumes commutativity and
associativity of addition and multiplication, and distributivity of multiplication with respect to
addition~\cite{Czachor_II,Naudts}. In consequence, whenever one encounters ``plus'' or ``times'' one has certain flexibility in
interpreting these operations.
A change of realization of arithmetic, without altering the remaining
structures of equations involved, plays an analogous role as a symmetry transformation in physics.

Let us considering a bijection $f^{-1}: M \mapsto N \subset
\mathbb{R}$, where $M$ is some set. The map $f$ allows us to define
addition, subtraction, multiplication, and division in $M$, as follows
\begin{eqnarray}
x\oplus y &=&  f(f^{-1}(x) + f^{-1}(y))\, ,\nonumber \\
x\ominus y &=&  f(f^{-1}(x) - f^{-1}(y))\, ,\nonumber \\
x\otimes y &=&  f(f^{-1}(x)f^{-1}(y))\, ,\nonumber \\
x\oslash y &=&  f(f^{-1}(x)/f^{-1}(y))\, .
\label{genar-a}
\end{eqnarray}
One can readily verify the following standard properties: (1) associativity
$(x\oplus y)\oplus z = x\oplus(y\oplus z)$, $(x\otimes y)\otimes z = x\otimes(y\otimes z)$,
(2) commutativity $x\oplus y = y \oplus x$, $x \otimes y = y \otimes x$, (3) distributivity
$(x \oplus y) \otimes z = (x \otimes z) \oplus (y \otimes z)$.  For a future convenience we will explicitly affiliate with the arithmetic operations $\oplus,\ominus,\otimes$ and $\oslash$ the symbol of the function $f$, so for instance, we will write $\oplus_f$ instead of $\oplus$, etc.

This generalized arithmetical structure motivated Kolmogorov and Nagumo~\cite{kolmogorov,nagumo} to
formulate the most general class of means,  so-called quasi-linear means, that are fully compatible with the usual Kolmogorov postulates of probability theory~\cite{Kolm-axioms}, with interesting applications in thermostatistics~\cite{Czachor}.

The aforemention generalized arithmetics can be extended quite naturally to real multivariate functions.
For instance, for a function of two variables $\mathcal{G}(x,y)$ it can be defined as
\begin{equation*}
\mathcal{G}_f(x,y) \ \equiv \  f\left(\mathcal{G}(f^{-1}(x),f^{-1}(y)\right)\, .
\end{equation*}
Let us state in this connection a couple of important consequences that can be easily verified:

\begin{enumerate}[label = {\em \roman*)}]
  \item when $z = x \otimes_f y$, then $g(z) = g(x) \otimes_{g \cdot f} g(y)$,
  \item $x \oplus_f y \ = \  x \otimes_{f \cdot \log}  y\, .$
Here, by $f \cdot g$ we implicitly mean the composition of two functions. Particularly important for our purposes will be the so-called $q$-deformed algebra where
\begin{eqnarray*}
~~~~~~~~f(x) \ \equiv \ f_q(x) \ =\  \log_q(x) \ = \ \frac{(x^{1-q}-1)}{(1-q)}\, .
\end{eqnarray*}
Ensuing operation $\otimes_{f_q}$ is traditionally denoted as $q$-addition and the notation $\oplus_q$ is often used instead.
  \item For the generalized product $\otimes_f$ the function $f$ is not determined uniquely. In fact, there exists a two-parametric class of functions $f_{a,b}$, so that
$f(x) \mapsto f_{a,b}(x) = f(ax^b)$, which yield the same product. Indeed,
\begin{eqnarray}
&&x \otimes_{f_{a,b}} y \nonumber \\
&&~~~~~= f\left(a \left[ (f^{-1}(x)/a)^{1/b} (f^{-1}(y)/a)^{1/b} \right]^b \right)\nonumber \\
&&~~~~~= x \otimes_f y\, .
 \label{II.7.a}
\end{eqnarray}
This result will be particularly important in Section~III.
\end{enumerate}

\section{Kolmogorov--Nagumo generalization of Shannon--Khinchin  axioms}

Let us now generalize the Shannon--Khinchin (SK) entropic axioms
in terms of the Kolmogorov--Nagumo arithmetics in the following way:

\begin{enumerate}[label = {\bf SK{\arabic*}}]
\item \emph{Continuity:} Entropy is a continuous function w.r.t. all its arguments, i.e.,
$S(P) \in \mathcal{C}.$
\item \emph{Maximality:} Entropy is maximal for uniform distribution, i.e.,
$\max_P S(P) = S(U_n)$,
where $U_n = \{1/n,\dots,1/n\}$.
\vspace{1mm}
\item \emph{Expandability:} Adding an elementary event with probability zero does not change the entropy, i.e,
\vspace{-2mm}
$$S(p_1,\dots,p_n,0) = S(p_1,\dots,p_n).$$
\vspace{-8mm}
\item \emph{Composability:} Joint entropy for a pair $(A,B)$ of random variables can be expressed as
\vspace{-2mm}
$$S(A,B) = S(A|B) \otimes_f S(B),$$
where $S(A|B)$ is conditional entropy satisfying consistency requirements {\bf I), II)} (see below).
\end{enumerate}
In passing, we can observe from the two illustrative axiomatic schemes ${\bf SK4}^R$ and ${\bf SK4}^T$ that viable entropic functionals should obey two natural conditions:

\begin{enumerate}[label = \bf \Roman*)]
  \item For two independent random variables $A$ and $B$ the joint entropy $S(A,B)$ should be \emph{composable} from entropies $S(A)$ and $S(B)$, i.e., $S(A,B) = F(S(A),S(B))$
  \item Conditional entropy should be \emph{decomposable} into entropies of conditional distributions, i.e., $S(B|A) = G\left( P_A, \{S(B|A=A_i)\}_{i=1}^n\right)$.
\end{enumerate}
Here $F$ and $G$ are functionals to be determined shortly. The motivation for these two conditions is taken from the original SK axioms for Shannon, R\'{e}nyi and Tsallis entropy. They all are \emph{composable} from marginal entropies if the subsystems are independent and they all are \emph{decomposable} into conditional entropies and (escort) marginal distributions.

Let us also note that the conditional entropy $S(A|B)$ automatically fulfills several important properties:
\begin{enumerate}[leftmargin=10mm,labelsep=2.9mm,label = {\alph*)}]
  \item \emph{Entropic Bayes' rule:} $S(A|B) \ = \ S(B|A) \ \oslash_f \ S(B) \ \otimes_f \ S(A)$ ,
  \item \emph{Generalized Gibbs inequality:} $S(A|B) \ \leq \ S(A)$.
\end{enumerate}
The Bayes rule is easy to show from the interchangeability of $S(A,B) = S(B,A)$ and by using the definition of conditional entropy. The second law of thermodynamics is easy to show because $S(A,B) \oslash_f S(B) \leq S(A)$.

Moreover, we can define the mutual information as
\begin{equation*}
I(A,B) \ = \  S(A,B) \oslash_f \left(S(B) \otimes_f S(A)\right)\, .
\end{equation*}
The composition requirement {\bf I)}  is  equivalent to $I(A,B) = f(1)$ for independent random variables.  We might note that the requirement {\bf I)} is equivalent to \emph{strict composability axiom} introduced in Ref.~\cite{tempesta}.

Let us now prove the following theorem:
\begin{theorem}
The most general class of entropic functionals $S$ satisfying the aforestated axioms \textbf{SK1-4} can be expressed as
\begin{equation}\label{eq:sk}
  S_q^f(P) \ = \ f\left[\left(\sum_i p_i^q\right)^{1/(1-q)}\right],
\end{equation}
where $f(x)$ is a generic strictly increasing function defined on $x\in [0,\infty)$.
\end{theorem}
In passing it is useful to note that (\ref{eq:sk}) can be equivalently expressed as
\begin{equation}\label{eq:sk2}
S_q^f(P) \ = \ f\left[ \exp_q \!\left(\sum_i p_i \log_q\left(\frac{1}{p_i}\right)\right)\right],
\end{equation}
where
%
$\exp_q(x) \ = \ [1+(1-q)x]^{1/(1-q)}$.
%
Eq.~(\ref{eq:sk2}) is a simple consequence of the fact that
\begin{eqnarray*}
 &&\exp_q \sum_i p_i \log_q\left(\frac{1}{p_i}\right) \nonumber\\ &&= \left(1 + (1-q) \sum_i p_i \frac{p_i^{1-q}-1}{1-q} \right)^{1/(1-q)}\nonumber\\
 &&= \left( \sum_i p_i^q  \right)^{1/(1-q)}\, .
\end{eqnarray*}

\noindent
\textbf{Proof of Theorem 1:} First, we see that the functional has to be symmetric in all components of $P=\{p_i\}$. This is because
by relabeling points in a set of elementary events should not change the information about the underlying stochastic process. Consequently, $S$ must be symmetric. Second, the entropy of the uninform distribution $S(n) \equiv S(1/n,\dots,1/n)$ can be obtained from composability axiom. To this end we denote the random variable with uniform distribution as $U_{nm} = U_n U_m$. We abbreviate $S(U_n)$ as $S(n)$. Then [see Eq.~(\ref{genar-a})]
\begin{equation*}
S(nm) \ = \  S(n) \otimes_f S(m) \Rightarrow S(n) \ = \ f(n)\, .
\end{equation*}
Here we have explored the freedom in the definition of the function $f$ [see, Eq.~(\ref{II.7.a})] and scaled back the generic  solution $S(n) = f(n^x)$, $x\in \mathbb{R}$ to the solution $S(n) = f(n)$.

Third, let us take two random variables $A$ and $B$ with distributions $P_A = \{p_i\}_{i=1}^n$ and $P_B =
\{q_j =1/m\}_{j=1}^m$. Let us also introduce the so-called Dar\'{o}czy mapping \cite{ilic,jizba17}, i.e., $S \mapsto f^{-1} S$. After this mapping we get multiplicative entropy. From the definition of $S(A|B)$ we then obtain that
\begin{eqnarray}
\label{eq:hom}
 m f^{-1}S(p_1/m,\dots,p_n/m) =  f^{-1}S(p_1,\dots,p_n) \, ,
\end{eqnarray}
since the conditional entropy is for each random variable just the usual unconditional one. Therefore, entropy must be a first order homogenous, symmetric function.
According to \cite{aczel89} the solution of homogeneous equation (\ref{eq:hom}) can be (under mildly restrictive assumptions)  expressed
as
\begin{equation}\label{eq:homogen}
f^{-1}S(x_1,\dots,x_n) = b \prod_{i=1}^n x_i^{a_i} \quad \mathrm{where} \ \sum_i a_i = 1\, .
\end{equation}
Here $a_i$ and $b$ are constants to be specified later. However, this solution is not symmetric in its variables. This can be achieved by symmetrization of Eq. \eqref{eq:homogen} that can be then rewritten in the following form
\begin{equation*}
f^{-1}S(p_1,\dots,p_n) \ = \ b \sum_{\{j_1,\dots,j_n\} \in \sigma(n)} \prod_{i=1}^n p_i^{a_{j_k}}\, .
\end{equation*}
This expression can be equivalently recast to
\begin{equation*}
f^{-1}S(p_1,\dots,p_n) \ = \  b \prod_{i=1}^n\left(\sum_{k_i} p_{k_i}^{a_{i}} \right)
\, ,
\end{equation*}
that can further be rewritten as
%
\begin{equation}
f^{-1}S(p_1,\dots,p_n) \ = \  b \prod_{i=1}^n\left(\sum_{k_i} p_{k_i}^{a_{i}} \right)^{c/(1-a_i)}
,
\label{III.15}
\end{equation}
which still keeps the entropy to be a homogeneous function of the first order. The parameter $c$ is a free parameter that will be determined later. Note that this representation is also mentioned in \cite{tempesta18}.

Let us now show that in order to fulfill the decomposability axiom \textbf{II)}, only one $a_{j}$ must be non-zero. To this end, we explicitly express $f^{-1}S(A|B)$ as
\begin{eqnarray*}
f^{-1} S(A|B) \ = \ \frac{ \left(\sum_{k_1,l_1} (r_{k_1|l_1}q_{l_1})^{a_{1}} \right)^{c/(1-a_1)}}{\left(\sum_{l_1} q_{l_1}^{a_{1}} \right)^{c/(1-a_1)}}\nonumber\\
\times \dots \times \frac{\left(\sum_{k_n,l_n} r_{k_n|l_n}q_{l_n}^{a_{n}}\right)^{c/(1-a_n)}}{\left(\sum_{l_n} q_{l_n}^{a_{n}}\right)^{c/(1-a_n)}}\, .
\end{eqnarray*}
This can be more explicitly rewritten as
\begin{eqnarray*}
\mbox{\hspace{-2mm}}f^{-1} S(A|B) \! = \!  \left(\sum_{l_1} \rho_{l_1}^{B}(a_1) \sum_{k_1} (r_{k_1|l_1})^{a_{1}} \right)^{c/(1-a_1)} \nonumber\\
\times \dots \times \!\left(\sum_{l_n} \rho_{l_n}^{B}(a_n) \sum_{k_n} (r_{k_n|l_n})^{a_{n}} \right)^{c/(1-a_n)}\!\!,
\end{eqnarray*}
where $\rho_{l}^B(a) = {q_l^a}/{\sum_l q_l^a}$ is the escort distribution~\cite{beck,Mendes}.
This expression is an unconditional entropy of the conditional distribution only
if one of $a_j$ is non-zero and the remaining ones are zero. With this we get that
\begin{eqnarray*}
\mbox{\hspace{-6mm}}f^{-1} S(A|B)\!\! &=& \!\! \left(\sum_{l} \rho_{l}^{B}(a) \sum_{k} (r_{k|l})^{a} \right)^{1/(1-a)} \nonumber \\ &=& \!\!\left\{\sum_{l} \rho_{l}^{B}(a) [S(A|B=b_l)]^{1-a}\right\}^{\!\!1/(1-a)}\!,
\end{eqnarray*}
which directly implies the decomposibility function $G$. With this result Eq.~(\ref{III.15}) boils down to
\begin{eqnarray*}
\mbox{\hspace{-6mm}}f^{-1}S(p_1,\dots,p_n) &=&  b \left(\sum_{k} p_{k}^{a} \right)^{c/(1-a)} \nonumber \\[1mm] &=&  b \left[\exp_a\left(\sum_{k} p_k  \log_a \frac{1}{p_k}\right)\right]^c,
\end{eqnarray*}
%
%
%
%
which by Eq.~(\ref{II.7.a}) is equivalent to (\ref{eq:sk}) provided we identify $a$ with $q$.

The function $f$ must be  strictly monotonic because in the proof we needed inverse of $f$, and must be strictly increasing because $S$ has by \textbf{SK2} the maximum for uniform distribution (and not, for instance, for $P = (1,0,0,\ldots ,0)$).
This  concludes the proof. \qed\\


Let us finally note that the original axiom {\bf SK4$^S$} is recovered from {\bf SK4} by taking $f(x) = \ln x$ and the decomposibility function $G(x_i,y_i) = \sum_i x_i y_i$.

\section{Equivalence with Shore--Johnson axioms}

Let us now turn our attention to MEP  and corresponding consistency requirements. The MEP  can be formulated in the following way~\cite{jaynes1,jaynes2}:\\

\noindent{\bf Proposition~}(Maximum entropy principle).~
{\em{Given the set of linear constraints $\sum_i p_i E_{i}^{(k)} = \langle E^{(k)} \rangle$, the least biased estimate of the underlying distribution $P=\{p_i\}$ is obtained from maximization of the entropic functional $S(P)$ under normalization constraint and set of constraints $\langle E^{(k)} \rangle$, i.e., by maximizing the Lagrange functional}}
\begin{equation}
S(P) - \alpha \sum_{i=1}^N p_i - \sum_{k=1}^\nu \beta^{(k)} \sum_{i=1}^N p_i E_i^{(k)}\  .
\label{22a}
\end{equation}
%

In (\ref{22a}) the index ``$i$'' runs over all possible  {\em states}, i.e., over all elements from the set of possible outcomes associated with a given random system.

Shore and Johnson formulated the set of consistency requirements that the MEP should satisfy~\cite{shore,shoreII}:
\begin{enumerate}[label = {\bf SJ{\arabic*}}]
\item \emph{Uniqueness}: the result should be unique.
\item \emph{Permutation invariance}: the permutation of states should not matter.
\item \emph{Subset independence}: It should not matter whether one treats disjoint subsets of system states in terms of separate conditional distributions or in terms of
the full distribution.
\item \emph{System independence}: It should not matter whether one accounts for independent constraints related to independent systems separately in terms of marginal distributions or in terms of full-system.
\item \emph{Maximality}: In absence of any prior information, the uniform distribution should be the solution.
\end{enumerate}

Let us now state without proof the theorem that provides the most general class of admissible
entropic functionals consistent with the aforestated~\textbf{SJ} axioms:
\begin{theorem}[Uffink theorem]
The class of entropic functionals $S$ satisfying the axioms \textbf{SJ1-5} can be expressed as
\begin{equation}\label{eq:sj}
  S_q^f(P) \ = \ f\left[\left(\sum_i p_i^q\right)^{1/(1-q)}\right],
\end{equation}
for any $q > 0$ and for any strictly increasing function $f$.

\end{theorem}
In particular, the Uffink theorem shows that members of this entropic class  admit representation in the form given by Eq.~(\ref{eq:sk}), and hence the  SK and SJ axiomatic systems are equivalent. A detailed proof of Theorem~2 can be found in Ref.~\cite{jizba19}.

Let us now discuss some salient results of the proof. First two axioms assert that the entropic functional must be a symmetric functional in the probability components. The third axiom determines the function in the sum form, i.e. in the form
%
$S(P) =f\left( \sum_{k} g(p_k)\right)$,
with $g$ being  an arbitrary increasing concave function.
The fourth axiom
gives us the final form of the entropic functional (without specifying the range for $q$'s), and finally the fifth axiom guaranties that $q>0$. Note that the class obtained from Theorem 1 and epitomized by Eq.~\eqref{eq:sk} is the same as the class given by
Eq.~\eqref{eq:sj} from Uffink theorem. Therefore, we immediately see
that in axiom {\bf SK4} the requirement {\bf II)} (decomposability) corresponds to axiom {\bf SJ3}, while requirement {\bf I)} (composability) corresponds to axiom {\bf SJ4}.
Moreover, the interpretation of $f$ and $q$ is now clear. The function $f$ determines the scaling of the entropy for uniform distribution (as it is independent of $q$), see also~\cite{korbel18}, while
the parameter $q$ determines the correlations in the system through MaxEnt distribution,
which can be expressed as (see Ref.~\cite{jizba19})
\begin{eqnarray*}
&&p_i \ = \  \frac{1}{Z_q} \exp_q\left(-\upbeta \Delta E_i \right)\, ,\nonumber \\[-1mm]
&& Z_q \ = \ \sum_i \left[\exp_q\left(-\upbeta \Delta E_i \right)\right]\, ,\nonumber \\[-1mm]
&&\upbeta \ = \ \frac{\beta}{q f'\left(Z_q\right) \, Z_q}\, ,
\end{eqnarray*}
%
%
where $ \Delta E_i = E_i - \langle E \rangle$. The connection of the $q$ parameter with correlations can be understood from the MaxEnt distribution of a joint system composed from two disjoint subsystems. Let us denote the MaxEnt distribution of the joint system as $p_{ij}$ and the MaxEnt distribution of the subsystems as $u_i$ and $v_j$. In~\cite{jizba19} it was shown that the MaxEnt distributions involved fulfill the composition rule that can be formulated as
\begin{equation}\label{eq:comp1}
\frac{1}{p_{ij} \, \mathcal{U}_q(P)} \  = \  \frac{1}{u_{i} \, \mathcal{U}_q(U)} \otimes_q \frac{1}{v_{j} \, \mathcal{U}_q(V)}\, .
\end{equation}
where $x \otimes_q y = [x^{1-q} + y^{1-q} - 1]_+^{1/(1-q)}$ (with $x,y > 0$) is the so-called $q$-product \cite{Borges04}, and $\mathcal{U}_q(P) = \left( \sum p_{ij}^q\right)^{1/(1-q)}$, and similarly for $U$ and $V$.
%
%
%
For $q \rightarrow 1$, (\ref{eq:comp1})  reduces to $p_{ij} = u_i v_j$.
The reverse is true as well. By re-expressing~\eqref{eq:comp1} in terms of escort distributions $P_{ij}(q)$, $U_i(q)$ and $V_j(q)$ (associated with $p_{ij}$, $u_i$ and $v_j$, respectively) as
\begin{equation}\label{eq:compes}
\frac{P_{ij}(q)}{p_{ij}} \ = \ \frac{U_{i}(q)}{u_{i}}\ + \ \frac{V_{j}(q)}{v_{j}} \ - \ 1\, ,
\end{equation}
and using $p_{ij} = u_i v_j$, we obtain $U_{i}(q) = u_{i}$,  $V_{k}(q) = v_{k}$ (for all $i,k$).
Latter have a unique solution~\cite{beck} $q=1$. This implies that $q$ parametrizes correlations
between system's subsystems since only for $q=1$ the Pearson correlation coefficient is zero.

As discussed, e.g., in~\cite{korbel19}, a monotonic function of an entropic functional gives the same MEP distribution and redefines only the Lagrange multipliers but does not change the actual form of the distribution. This can be interpreted as a sort of \emph{gauge invariance} $S(P) \mapsto f(S(P))$. Finally, let us mention that the $q=1$ case corresponds to uncorrelated MEP distributions for disjoint systems, for which we get a stronger version of system independence axiom~\cite{jizba19}:
\begin{description}
  \item[{\bf SJ4}$^{SSI}$] \emph{Strong system independence:} Whenever two subsystems of a system are disjoint, we can treat the
subsystems in terms of independent distributions.
\end{description}
%
The solution is then
\begin{equation*}
S_1^f(P) = f\left(\exp\left[\sum_i p_i \log\left(1/p_i\right)\right]\right)\, ,
\end{equation*}
which is equivalent (through Dar\'{o}czy mapping) to Shannon entropy --- as expected. In this case, the composition rules in Eqs.~\eqref{eq:comp1} and~\eqref{eq:compes} reduce to the composition rule of independent systems, i.e.,
\begin{equation*}
p_{ij} \ = \  u_i v_j\, .
\end{equation*}

On the other hand,
if we require that the entropy must be in the trace form~\cite{tempesta,hanel}, i.e., $S(P) = \sum_i g(p_i)$, then we get that $f(x) = \log_q(x)$ and we end up with the class of THC entropies
\begin{equation*}
S_q^{\log_q}(P) \ = \  \sum_i p_i \log_q\left(1/p_i\right)\, .
\end{equation*}

\section{Conclusions}

Here we have reformulated Shannon--Khinchin axioms of information theory in terms of generalized arithmetics of Kolmogorov and Nagumo. Apart from the axiomatic structure itself, the novelty of this work is in showing that the general class of entropic functional satisfying such SK axioms is identical with the Uffink class of entropies. Since the Uffink class is known to represent the general solution of Shore--Johnson axioms of statistical-inference theory, both axiomatic systems have to be equivalent. We have shown that Uffink functionals $S_q^f$ are characterized by the Kolmogorov--Nagumo function $f$ and a positive parameter $q$, where $f$ determines a scaling behavior of entropy for uniform distributions and $q$ quantifies correlations of MEP distributions for disjoint subsystems. In passing, we can note that the form (\ref{eq:sk2}) of the class $S_q^f$ can also be found in the literature under the name \emph{strongly pseudo-additive} (SPA) entropies~\cite{ilic19} or $Z$-entropies~\cite{tempesta}.


The outlined entropic parallelism between information theory and statistical inference can serve as a good starting point for further research. In this context it
would be particularly interesting to investigate how robust the aforementioned equivalence between the two axiomatic systems is and assess the extent and consequences resulting from a prospective breakdown. One might instigate such a breakdown by working, e.g. with more general constraints (non-linear constraints or scalings as in non-inductive inference) or by relaxing some of the presented axioms. In fact, it is well-known that many complex systems do not satisfy SK axioms, not even in our generalized sense~\cite{cht15,hanel,korbel18}. By relaxing some of these axioms, one might gain further maneuvering space allowing to accommodate entropies of such systems as path-dependent and super-exponential systems or complex systems with non-trivial constraints.

\section*{Acknowledgements}
P.J. and J.K. were supported by the Czech Science Foundation (GA\v{C}R), Grant 19-16066S. J.K. was also supported by the Austrian Science Foundation (FWF) under project I3073.

\end{document}